\documentclass[fleqn,usenatbib]{mnras}

\usepackage[T1]{fontenc}
\usepackage{ae,aecompl}

\usepackage{graphicx}	% Including figure files
\usepackage{amsmath}	% Advanced maths commands
\usepackage{amssymb}	% Extra maths symbols

\title[Gravitational millilensing in B1152+199]{The case for gravitational millilensing in the multiply--imaged quasar B1152+199}

\author[Asadi et al.]{Saghar Asadi$^{1}$\thanks{E--mail: saghar.asadi@astro.su.se}, Erik Zackrisson$^{2}$, Eskil Varenius$^{3,5}$, Emily Freeland$^{1}$, John Conway$^{3}$, \newauthor Kaj Wiik$^{4}$
\\
$^{1}$Department of Astronomy, Stockholm University, Oscar Klein Center, AlbaNova, Stockholm SE--106 91, Sweden\\
$^{2}$ Department of Physics and Astronomy, Uppsala University, Box 515, SE--751 20 Uppsala, Sweden\\
$^{3}$ Department of Earth and Space Sciences, Chalmers University of Technology, Onsala Space Observatory, 439 92 Onsala, Sweden\\
$^{4}$ Tuorla Observatory, Department of Physics and Astronomy, University of Turku, V\"ais\"al\"antie 20, FI--215 00 Piikki\"o, Finland\\
$^{5}$ Jodrell Bank Centre for Astrophysics, The University of Manchester, Oxford Rd, Manchester M13 9PL, UK.\\
}

\date{Accepted XXX. Received YYY; in original form ZZZ}

\pubyear{2018}

\begin{document}
\label{firstpage}
\pagerange{\pageref{firstpage}--\pageref{lastpage}}
\maketitle

% Abstract of the paper (max 250 words - currently 220)
\begin{abstract}
Previous Very Long Baseline Interferometry (VLBI) observations of the quasar B1152+199 at 5\,GHz has revealed two images of a strongly lensed jet with seemingly discordant morphologies. Whereas the jet appears straight in one of the images, the other exhibits slight curvature on milliarcsecond scales. This is unexpected from the lensing solution and has been interpreted as possible evidence for secondary, small--scale lensing (millilensing) by a compact object with a mass of $~10^5$--$10^7\ M_\odot$ located close to the curved image. The probability for such a superposition is extremely low unless the millilens population has very high surface number density. Here, we revisit the case for millilensing in B1152+199 by combining new global--VLBI data at 8.4\,GHz with two datasets from the European VLBI Network (EVN) at 5\,GHz (archival) and at 22\,GHz (new dataset), and the previously published 5\,GHz Very Long Baseline Array (VLBA) data.

We find that the new data with a more circular synthesized beam, exhibits no apparent milliarcsecond--scale curvature in image B. Various observations of the object spanning $\sim$15 years apart enable us to improve the constraints on lens system (thanks also to the improved astrometry resulting from 22\:GHz observations) to the point that the only plausible explanation left for the apparent curvature is the artifact due to the shape of the synthesized beam.
\end{abstract}

\begin{keywords}
Gravitational lensing: strong -- dark matter -- quasars -- galaxies: jets 
\end{keywords}

%%%%%%%%%%%%%%%%%%%%%%%%%%%%%%%%%%%%%%%%%%%%%%%%%%

%%%%%%%%%%%%%%%%% BODY OF PAPER %%%%%%%%%%%%%%%%%%

\section{Introduction}
A generic prediction of the the cold dark matter (CDM) model is the existence of dark halo substructure (a.k.a. subhalos) in the mass range of dwarf galaxies and below. The mismatch between the observed luminosity function of satellite galaxies and the predicted mass function of dark matter halos also implies that the vast majority of these substructures need to be extremely faint or completely dark at optical wavelengths. Pinning down the properties of this dark subhalo population provides an important test between CDM and alternative dark matter models which often predict subhalos with significantly different mass functions and density profiles \citep[e.g.][]{Li16,Bose16}. 

Methods for hunting down dark matter subhalos include the search for dark matter annihilation signals \citep[e.g.][]{Schoonenberg16,Hutten16,Mirabal16}, the perturbations of gas, stars and stellar streams in the Milky Way \citep[e.g.][]{Chakrabarti11,Feldmann15,Carlberg16,Erkal16}, HI clouds with no stellar counterpart \citep[e.g.][]{Keenan16} and effects of small--scale gravitational lensing. In the latter case, the main targets are distant galaxies and quasars that are strongly lensed by a foreground galaxy. By looking for small lensing perturbations introduced by substructure in the halo of the main lens, even completely dark subhalos can in principle be detected \citep[see][for a review]{Zackrisson10}. Using optical, near--infrared and sub--millimeter observations, this technique has already revealed various kind of lens substructure in the dwarf galaxy mass range \citep[e.g.][]{Vegetti10,MacLeod13,Nierenberg14,Inoue16,Hezaveh16}, and at least one strong case for a dark or extremely faint subhalo \citep[][]{Vegetti12}. 

Very long baseline interferometry at radio or sub--mm wavelengths is currently the only technique that can detect such perturbations at milliarcseconds scale and below \citep[e.g.][]{Wilkinson01,Zackrisson13,Fish13}, but the small intrinsic sizes of the sources that are sufficiently bright for current observations of this type severely limits the categories of substructures that can be detected through such millilensing effects. Basically, only halo objects much denser than predicted in the vanilla cold dark matter scenario can be detected this way \citep{Zackrisson13} -- like ultracompact minihalos and primordial black holes. 

Curiously, VLBI observations have already produced one tentative detection of gravitational millilensing -- in the B1152+199 strong-lensing system, first identified by the Cosmic Lens All-Sky Survey (CLASS, \citealt{Myers99}). In this case, a quasar at $z=1.0189$ is lensed into two images A and B by a galaxy at $z=0.4386$. VLBA observations by \citet{Rusin02}
have revealed that the two images, which are about 1.56 arcsec apart, exhibit a jet-core structure with a flux ratio of $\sim$\,3:1. Optical observations by \citet{Toft00} indicate that image B -- which is less magnified and passes closer to the lens galaxy -- is subject to considerable dust attenuation, corresponding to a differential reddening of $E(B-V)\approx 1.0\pm 0.1$ mag. The time delay between the images is estimated at $\approx 30$--60 days \citep{Toft00,Edwards01,Munoz01,Rusin02}.

As first noted by \citet{Rusin02}, the jet of image B exhibits a slight curvature on milliarcsecond scales that is not seen in the image A jet. This was interpreted by \citet{Metcalf02} as millilensing of an object of mass $\sim 10^5$--$10^7\ M_\odot$ in the vicinity of image B. However, \citet{Metcalf02} also find that the type of substructure required to explain this feature would need to belong to a very numerous population of objects. Luminous objects in this mass range, like globular clusters and dwarf galaxies, fall short by a large factor. If the object is a dark matter halo or subhalo, it would also need to be far more compact than predicted by the standard CDM.

Here, we present new and archival VLBI observations of the B1152+199 system to revisit the millilensing interpretation. The VLBI data is described in Section~\ref{section:data}, the overall properties of the B1152+199 lensing system in Section~\ref{section:modeling} and the case for jet curvature in image B due to gravitational millilensing in Section~\ref{section:discussion} where we also discuss the millilensing interpretation in the light of the current constraints on the nature of dark matter. Our findings are summarized in Section~\ref{section:conclusions}.

\section{Observational data}
\label{section:data}

\begin{table*}
\centering
\caption{Summary of the observational data sets considered for the B1152+199 core--jet system}
\label{table:datasets}
\begin{tabular}{lllll}
                                 & VLBA                 & EVN 5\,GHz          & EVN 22\,GHz         & global VLBI             \\
\hline\\
 Project Number                  & BB0133               & EJ010               & EZ024               & GA036             \\
 PI                              & Biggs                & Jackson             & Zackrisson          & Asadi             \\
 Epoch                           & Feb.--Mar. 2001      & Feb. 2012           & Feb. 2013           & June 2015         \\
 Observing frequency [GHz]       & 5                    & 5                   & 22                  & 8.4               \\
 beam size [mas]                 & 3.3 $\times$ 1.6     & 2.3 $\times$ 1.8    & 0.9 $\times$ 0.4    & 1.2 $\times$ 0.5  \\
 beam position angle [degrees]   & -9.2                 & 0.4                 & 2.1                 & -6.8              \\
 RMS noise  [$\mu$Jy/beam]       & 63                   & 17                  & 160                 & 26              \\
\end{tabular}
\end{table*}

In this paper we analyze four VLBI observations of B1152+199, summarized in Table \ref{table:datasets}, including the 5\,GHz VLBA data discussed by \citet{Rusin02} and \citet{Metcalf02}. All datasets were calibrated and imaged from the raw archival data using the same software --AIPS 31DEC15 \citep{greisen2003} and ParselTongue 2.3 \citep{kettenis2006}-- to reduce systematics due to different software and reduction strategies. In this section we describe the calibration and imaging of these four epochs.  

\subsection{BB0133: 5\,GHz VLBA}
\label{subsection:BB0133}
The VLBA program BB0133 was observed on two days: February 27 and March 18, 2001. One single correlation center was used between the two lensed images. These data were first presented in \citet{Rusin02}. Bandpass corrections were derived towards 3C279 and the visibility phases were referenced to J1148+1840, assuming a position from the RFC2016c calibrator catalog of R.A.$11^h48^m37.776758^s$, Dec.$18^\circ40'08.96888''$ at 1.9$^\circ$ separation from the target field. After transferring the cumulative corrections the target was imaged. Prominent phase errors were detected, such as a double structure due to systematic phase-differences between the two observing days, even after phase-referencing. We cleaned the image assuming a clean box around the brightest peak in the A-image, and performed one round of phase-only self-calibration using the cleaned image as input model. A new image was made, and another round of self-calibration was performed, now solving also for amplitudes to adjust minor antenna offsets in particular at the start and end of the observing sessions. The corrections
derived, once every minute, were inspected visually and found to vary slowly as expected. The cumulative corrections were applied to the target and the two lens images were imaged using \cite{briggs1995} weighting scheme with robustness parameter 0.5.  We note that although the two observations in this experiment both used the same phase-reference calibrator, clear phase-errors were found towards the target after applying the calibrator solutions. Hence, although the relative antenna errors were corrected for by hybrid imaging of the target to provide phase coherence through out the data, these data should not be used for absolute astrometry of the target field.

\subsection{EJ010: 5\,GHz EVN}
\label{subsection:EJ010}
The EVN program EJ010 was observed with 12 EVN telescopes and a single correlation center was used between the two lensed images.  Bandpass corrections were derived towards 3C345 and the visibility phases were referenced to J1143+1834, assuming a position from the RFC2016c calibrator catalog of R.A. $11^h43^m26.069659^s$, Dec.$18^\circ34'38.36006''$ at 3.0$^\circ$ separation from the target field.  After transferring the cumulative corrections the target was imaged. The target image was found to be dynamic range limited due to phase errors. To correct these, multiple rounds of phase self-calibration were performed followed by one round of amplitude self-calibration. A box-cleaned phase-referenced image was used as staring model and solutions were found once per minute. Particular care was taken to antenna SH which did not have enough signal to noise for reliable amplitude corrections to be found.  The cumulative corrections were applied to the target and the two lens images were imaged using \cite{briggs1995} weighting scheme with robustness parameter 0.5.  

\subsection{GA036: 8.4\,GHz Global VLBI}
\label{subsection:GA036}
The Global VLBI program GA036 was observed with 22 telescopes, including the VLBA, and used separate correlation centers for the two lens images.  Bandpass corrections were derived towards J1224+2122 and the visibility phases were referenced to J1158+1821, assuming a position from the RFC2016c calibrator catalog of R.A.$11^h58^m20.284711^s$, Dec.$18^\circ21'49.25433''$ at 1.5$^\circ$ separation from the target field.  After transferring the cumulative corrections the target was imaged. The target image was found to be dynamic-range limited due
to phase errors.  To correct these, multiple rounds of phase only self-calibration were performed (but no amplitude self-calibration) using a
solution interval of 2 minutes. The brightest A-image was used to derive the corrections, using the phase-referenced image as starting model. The cumulative corrections were applied to both correlated data sets (e.g. both lens images). Both images were imaged using \cite{briggs1995} weighting scheme with robustness parameter 0.5. We note that although we did not manage to remove all residual phase errors from the B-image, we think that these data are the best for absolute astrometry as they have the smallest target-phase reference calibrator separation.

\subsection{EZ024: 22\,GHz EVN}
\label{subsection:EZ024}
The EVN program EZ024 was rather unlucky. Of 14 scheduled telescopes, only 10 show fringes towards the phase-calibrator. Furthermore, multiple antennas (of the 10 with fringes) only recorded very little useful data as they failed for significant time and/or frequency ranges. Therefore, these data are of limited usability and should be treated with caution. Still, we managed to obtain images of the two lens images, using a similar calibration strategy as for the other data sets.  Two correlation centers were used (towards the two lens images).  Bandpass corrections were derived towards J1125+2610 and the visibility phases were referenced to J1157+1638, assuming a position from the RFC2016c
calibrator catalog of R.A.  $11^h57^m34.8362630^s$, Dec.$16^\circ38'59.649870''$ at 3.1$^\circ$ separation from the target field. The calibration procedure closely matches that described for the 8.4\,GHz data. Except that the phase-only self-calibrations were performed using a solution interval of 1 minutes to track the relatively rapid phase variations possible at 22\,GHz. The brightest A-image was used to derive the corrections, using the phase-referenced image as starting model. Given the limited number of telescopes and relatively large angular separation between calibrator and target, these data are of limited use due to the relatively poor sensitivity and astrometry. However, they do show that the luminous cores of both lens images are compact even at the high resolution obtained at 22\,GHz.

\begin{table*}
\centering 
\caption{Measured coordinates in units of arcseconds for the core and jet (outermost blob) of B1152+199, setting $A_\mathrm{core}$ as the origin. Errors on all positions are $\approx$ 1 mas.} 
\label{table:coordinates} 
    \begin{tabular}{|l|c|c|c|c|}
        \hline
        ~                 & VLBA (2001)        & EVN (2012)         & EVN (2013)         & global VLBI (2015)       \\
        ~                 & $x_1$\: $x_2$      & $x_1$\: $x_2$      & $x_1$\: $x_2$      & $x_1$\: $x_2$            \\ \hline
        $A_\mathrm{jet}$  & 0.0153\: $+$0.0011 & 0.0168\: $+$0.0009 & ~                  & 0.0179\: $+$0.0099       \\ 
        $B_\mathrm{core}$ & 0.9353\: $-$1.2454 & 0.9354\: $-$1.2455 & 0.9354\: $-$1.2455 & 0.9354\: $-$1.2455       \\ 
        $B_\mathrm{jet}$  & 0.9308\: $-$1.2497 & 0.9310\: $-$1.2495 & ~                  & 0.9306\: $-$1.2403       \\
        \hline
    \end{tabular}
\end{table*}

\begin{figure*} 
\includegraphics[width=\textwidth]{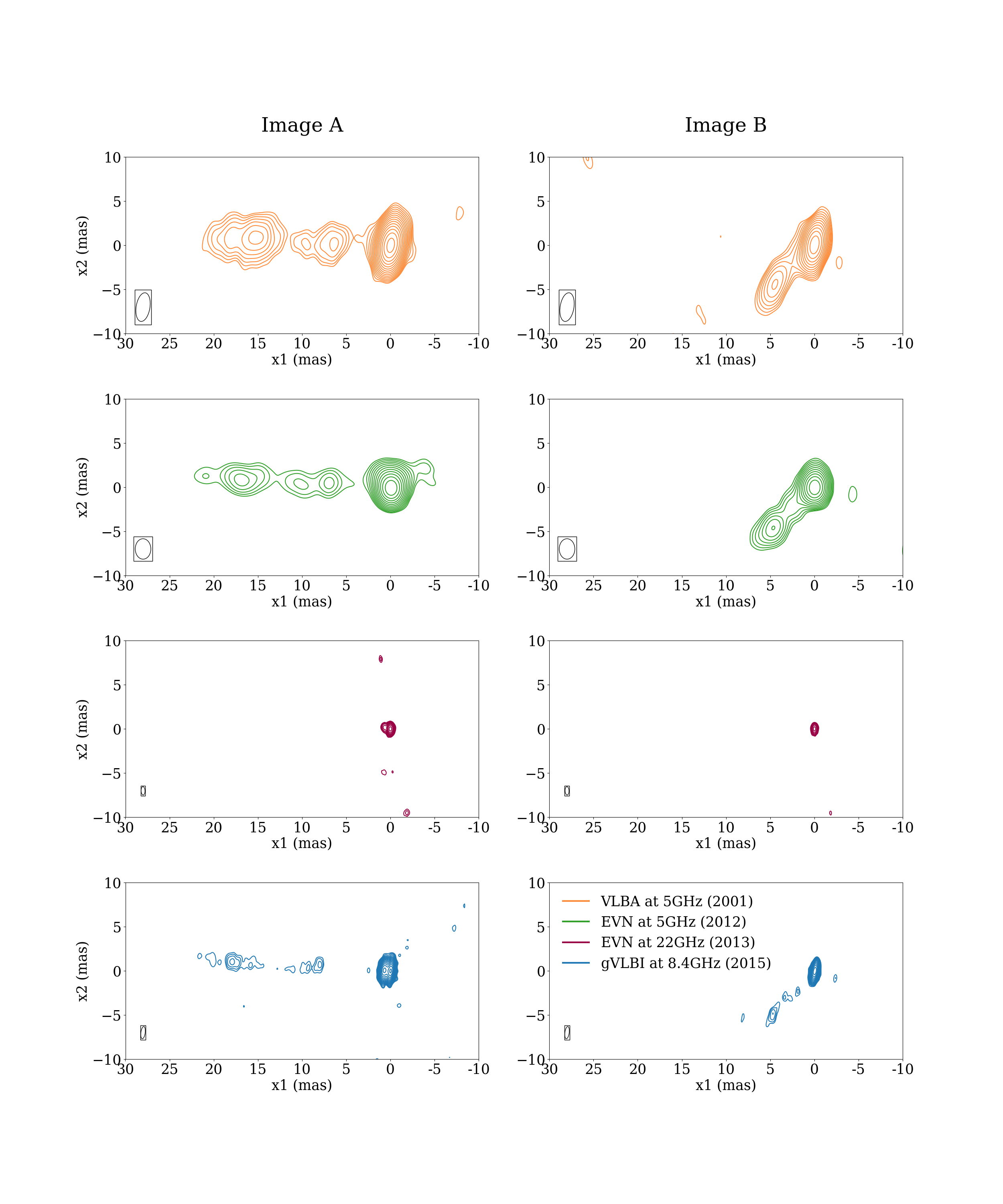} 
\caption{The two macrolensed images of B1152+199 from the re-imaged 1999 VLBA 5\,GHz data set (first row from the top), the 2012 EVN 5\,GHz data set (second row from the top), the 2013 EVN 22\,GHz data (third row from the top), and the 2015 Global VLBI array 8.4\,GHz data set (bottom row). In each case, the beam size is indicated in the lower left corner. The faintest contours are set to 3, 3, 5, and 10 times the rms noise respectively from top to bottom. Each subsequent contour is set at $\sqrt{2}$ times the previous one.}
\label{figure:jet_images} 
\end{figure*} 

\subsection{Images of the macrolensed jet}
\label{subsection:images_of_the_macrolensed_jets}
In Figure~\ref{figure:jet_images}, we present the contour maps of images A and B (lowest contour level is set to 3, 10, and 3 times the rms noise for the rows respectively from top to bottom and each level is a factor of $\sqrt{2}$ times the previous level) from the three data sets described in Sections~\ref{subsection:BB0133},~\ref{subsection:EJ010}, ~\ref{subsection:EZ024}, and~\ref{subsection:GA036}.

The apparent length of the jet is comparable in both 5\,GHz data sets and in the 8.4\,GHz set ($\approx 20$ mas for image A; $\approx 8$ mas for image B). While the jet in image A appears continuous in the VLBA 5\,GHz data, it breaks up into separate blobs in the other data sets, likely due to combination of different image fidelity, sensitivity limits and movement of blobs along the jet between the observation epochs. 
In agreement with \citet{Rusin02}, we find that image A appears straight with the best--fit position angle varying no more than $\approx 1\deg$ between the data sets (see Table~\ref{table:datasets}), once the small blob protruding from the core in the EVN 5\,GHz map is ignored. This weak western blob extension could possibly be a hint of a counter--jet feature, but could also be an artifact due to e.g. phase residuals in the data. We ignore this feature in our analysis and focus on the clear eastern jet structure.

The flux ratios between the cores of image A and B are $\approx 3$ to 1 in the two 5\,GHz data sets, but closer to $\approx 4.5$ to 1 in the 8.4\,GHz data set. This is unexpected from a lensing point of view, since the flux ratio is set by the macrolens solution and is not expected to change over time scales of decades. However, the core of image A in the 8.4\,GHz data set has double peaks, which may suggest that a very bright blob is currently emerging and therefore temporarily boosting the apparent flux of image A.

The position angle of image B is clearly different from the jet in image A, but this is -- as explained in Section~\ref{subsection:macrolens} -- expected from the strong lensing (macrolensing) by the main lens galaxy. The tantalizing aspect of image B, on which the case for potential millilensing by a massive object in the vicinity of this object is based, is that in two out of three datasets the jet exhibits slight downward {\it curvature} -- i.e. a trajectory that differs from a straight line. This curious feature is most clearly seen in the VLBA data set, and a hint of this is also present in the global VLBI maps, even though it is -- on its own -- less convincing due to the gaps between the jet blobs.

\section{The case for jet curvature in image B}
\label{section:modeling}

\subsection{The macrolens model}
\label{subsection:macrolens}
To investigate the expected shape of image B given a macrolens model without substructure, we use the Glafic lensing code \citep{Oguri10} to model the B1152+199 main lens as a singular isothermal ellipsoid (SIE) with external shear. Constraints on the best--fitting solution are set by the observed core and jet positions of image A, the core of image B, the flux ratio between the images, the position of the center of the SIE main lens, its position angle and ellipticity. We consider two options for the center of the external shear component -- that is either allowed to vary freely (referred to as the {\it free} model hereafter), or is forced to match the position of the main SIE (the so--called {\it match} model). The position and position angle of the main lens are assumed to lie within some interval from the values inferred by \citet{Rusin02} for the lens galaxy detected in Hubble Space Telescope (HST) $I$--band images. An 0.1 arcsecond offset between the SIE center and the the inferred center of that galaxy is considered acceptable, and so is a SIE position angle in the $-180\deg$ to  $0\deg$ range (for comparison, the value favored by \citealt{Rusin02} is $\approx -70\deg$).

\begin{figure*}
\includegraphics[width=\columnwidth]{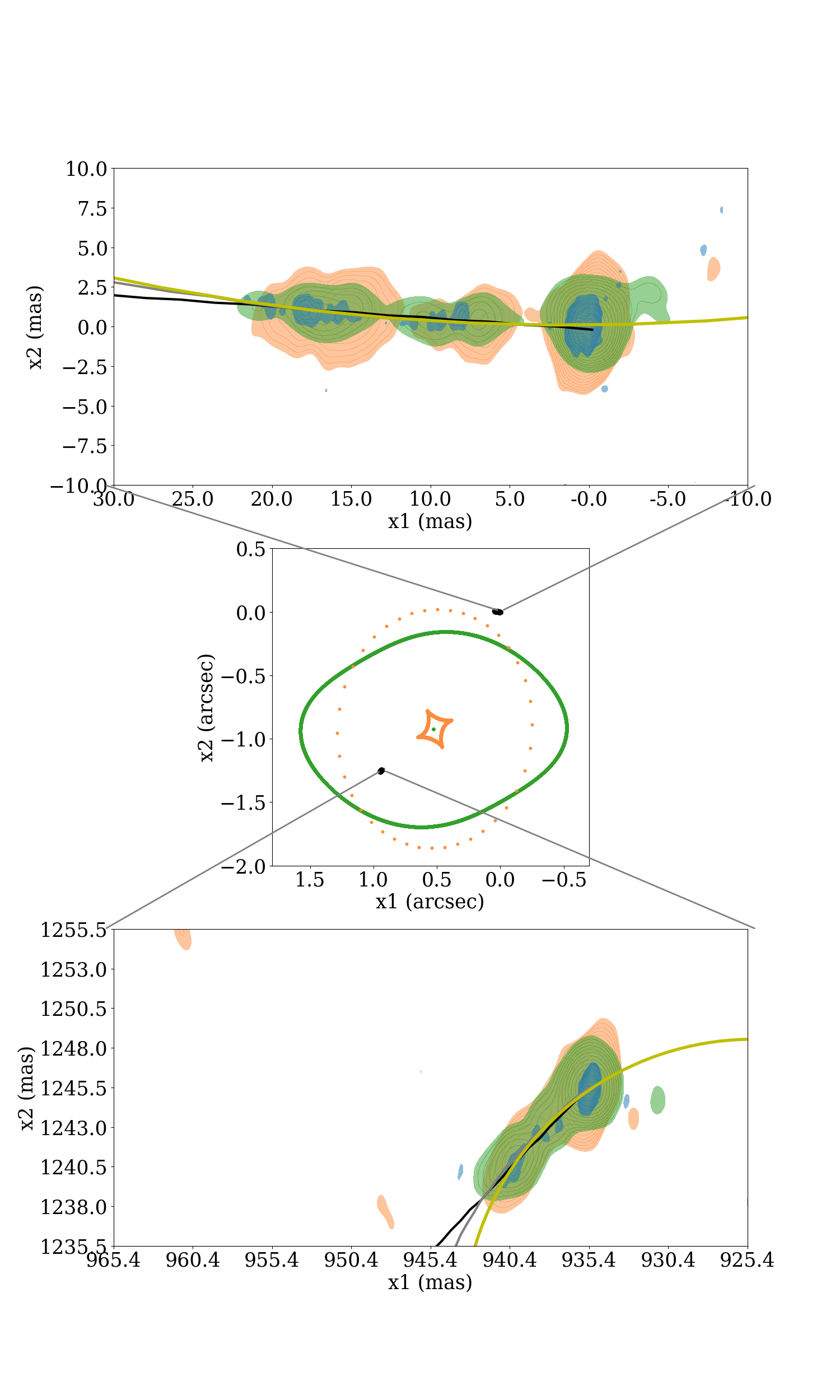}
\includegraphics[width=\columnwidth]{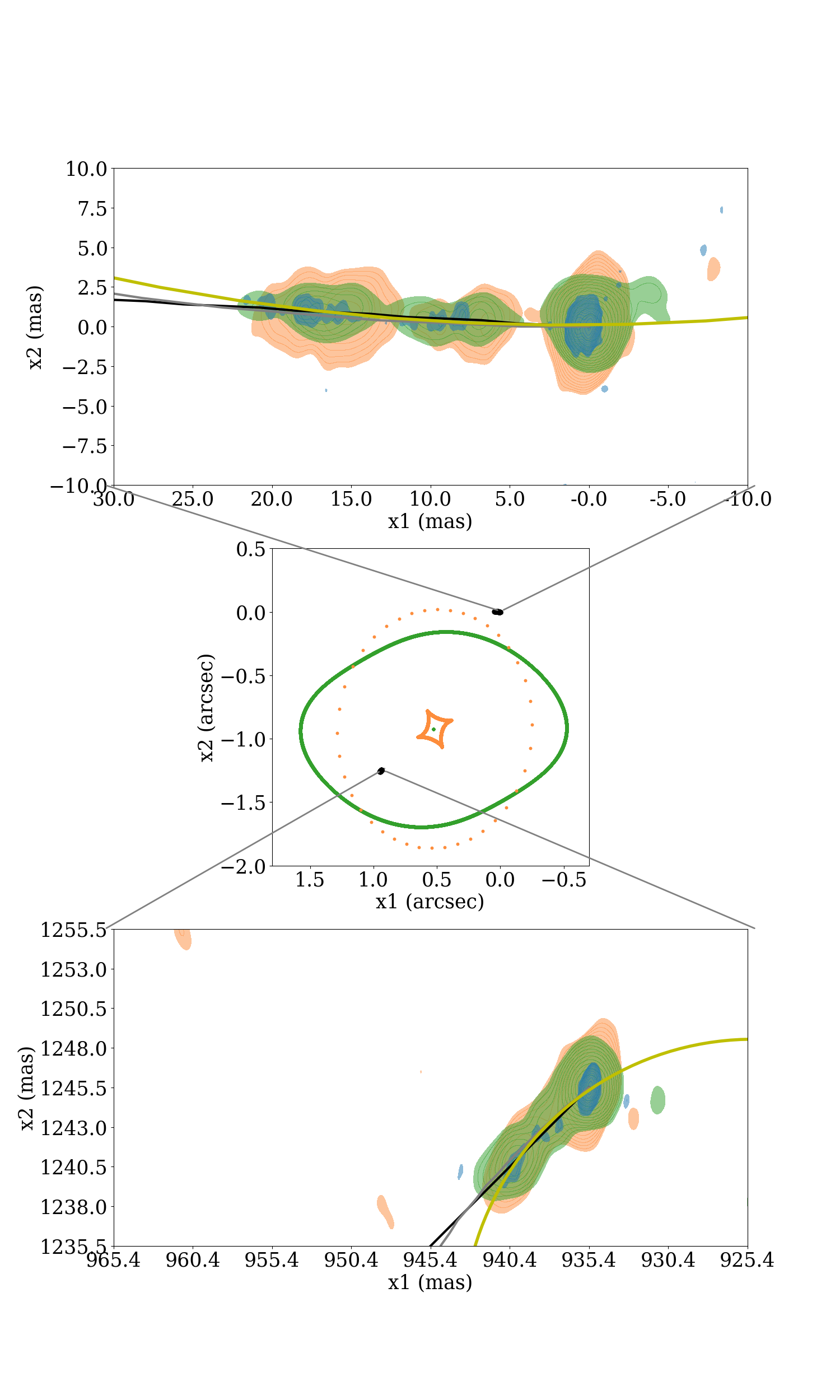}  
\caption{The macrolens solutions for the combined data sets of the B1152+199 system as described in Table~\ref{table:model_fit}. The left panel shows the {\it free} model and the right panel corresponds to the best--fitting {\it match} model to the combination of data sets. The critical and caustic curves of each lens model are drawn in green and orange, respectively. The upper and lower panels of each column are the zoomed--in regions around the macroimages with the predicted image positions overlaid the observational data. The contour plots follow the color scheme and levels as those in Figure~\ref{figure:jet_images}. Solid black lines correspond to an intrinsically straight jet, while the gray lines show how an intrinsically curved jet would deviate from the straight jet. The yellow lines in each panel show the best--fitting curvature of the corresponding image based on the measured peak positions. Note that the yellow line only serves to illustrate the apparent curvature of image B beyond what can be expected from intrinsic jet curvature (gray line).}
\label{figure:macrolens_solution} 
\end{figure*}

This procedure is performed for each of the three data sets (all datasets included in Figure \ref{figure:jet_images} but EVN at 22\: GHz) separately, but also for the combined map based on  centering all maps on the core position of image A (see Table~\ref{table:coordinates} for relative position measurements in this coordinate system). In Figure~\ref{figure:macrolens_solution}, we present the large--scale positions of the B1152+199 images with respect to the best--fit solution for this combined data set. The corresponding model parameter values are presented in Table~\ref{table:model_fit}, with uncertainties given by the range of best--fit values produced by the different fits to the individual data sets. All these best--fitting solutions have reduced $\chi^2_\mathrm{red}\approx 2$ and correspond to largely consistent macrolens parameter values.

\begin{figure*}
\includegraphics[width=\textwidth]{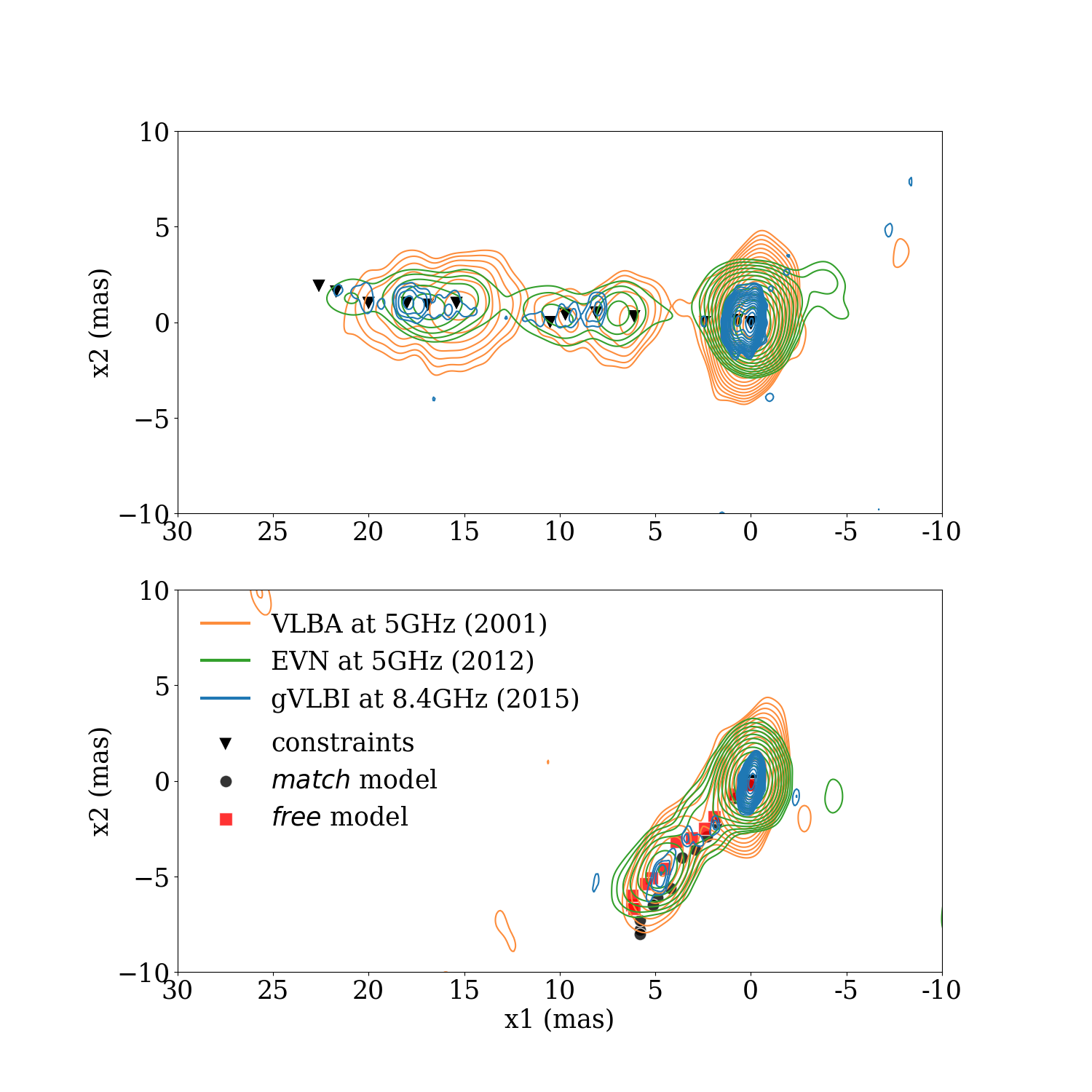}
\caption{Overlaid contour maps from the 5\,GHz VLBA observations (orange), 5\,GHz EVN observation (green) and 8.4\,GHz global VLBI observations (blue) of image A (upper row) and image B (lower row) along with model fits (image A jet) and predictions (image B jet) assuming a macrolens with no substructure. The core positions of both images, and the brightest jet blobs of image A are used to constrain the model and predict the positions of the corresponding blobs in image B. The black circles ({\it match} model) and red boxes ({\it free} model) depict the predicted positions of the corresponding peaks (black triangles) of image A, in image B.}
\label{figure:curvature} 
\end{figure*}

Our fitting procedure differs in its assumptions from that carried out by \citet{Rusin02} in the adopted main lens density profile. Whereas \citet{Rusin02} consider power--law density profiles with a few different slopes for the main lens, we have fixed the main lens to be of SIE type. In general, our inferred best--fit lens parameters are similar to theirs, with one important difference. Whereas the observed position angle of the jet in image B reported by them is not completely recovered by their fitting procedure, we have no problem reproducing the broad direction of this jet, as discussed in Section~\ref{subsection:curvature} and as shown on the image B in Figure~\ref{figure:curvature}. This is most likely due to updates in the astrometry of the two macroimages and to differences in how the position angle of the jet in image B is measured from the maps (not unambiguous due to the apparent curvature).

%% V3
\begin{table*}
  \centering 
  \caption{Best--fit parameters for the macrolens model of B1152+199, using the combined measurements of all three data sets} 
  \label{table:model_fit} 
    \begin{tabular}{llllll}
      {\it match} model & ~ & ~ & ~ & ~ & ~ \\
      \hline \hline
      ~ &  {\bf SIE $\sigma$ (km/s)} & {\bf SIE x$_1$ (arcsec)} & {\bf SIE x$_2$ (arcsec)} & {\bf SIE e} & {\bf SIE PA (deg)} \\
      \hline
      ~ & 2.46$^{+0.02}_{-0.09}\times$10$^2$ & 0.52$^{+0.05}_{-0.02}$ & -0.93$^{+0.08}_{-0.03}$ & 0.43$^{+0.01}_{-0.02}$ & -89$^{+6}_{-8}$ \\ \\
      ~ & {\bf dt (days)} & {\bf shear x$_1$ (arcsec)} & {\bf shear x$_2$ (arcsec)} & {\bf shear $\gamma$} & {\bf shear PA (deg)} \\
      \hline
      ~ &  31$^{+8}_{-2}$  & 0.50$^{+0.05}_{-0.02}$ & -0.97$^{+0.06}_{-0.08}$ & 0.15$^{+0.03}_{-0.02}$ & -165$^{+51}_{-20}$ \\ \\
      ~ & {\bf core x$_1$ (arcsec)} & {\bf core x$_2$ (arcsec)} & ~ & ~ & ~  \\
      \hline
      ~ & 0.33$^{+0.06}_{-0.04}$ & -0.68$^{+0.06}_{-0.06}$ & ~ & ~ & ~ \\ \\
      {\it free} model & ~ & ~ & ~ & ~ & ~ \\
      \hline \hline
      ~  &  {\bf SIE $\sigma$ (km/s)} & {\bf SIE x$_1$ (arcsec)} & {\bf SIE x$_2$ (arcsec)} & {\bf SIE e} & {\bf SIE PA (deg)} \\
       \hline
      ~  & 2.46$^{+0.04}_{-0.06}\times$10$^2$ & 0.55$^{+0.04}_{-0.03}$ & -0.90$^{+0.06}_{-0.03}$ & 0.39$^{+0.02}_{-0.02}$ & -80$^{+9}_{-12}$ \\ \\ \hline
      ~ & {\bf dt (days)} & {\bf shear x$_1$ (arcsec)} & {\bf shear x$_2$ (arcsec)} & {\bf shear $\gamma$} & {\bf shear PA (deg)} \\
      \hline
      ~ & 31$^{+6}_{-5}$ & 0.55$^{+0.03}_{-0.06}$ & -0.90$^{+0.08}_{-0.04}$ & 0.16$^{+0.02}_{-0.09}$ & -159$^{+43}_{-23}$ \\ \\
      ~ & {\bf core x$_1$ (arcsec)} & {\bf core x$_2$ (arcsec)} & ~ & ~ & ~  \\
      \hline
      ~ & 0.37$^{+0.04}_{-0.05}$ & -0.65$^{+0.07}_{-0.08}$ & ~ & ~ & ~ \\ 
\hline
    \end{tabular}
\end{table*}

\subsection{Jet curvature}
\label{subsection:curvature}
Figure~\ref{figure:macrolens_solution} depicts the image positions and critical curves of both the {\it free} and {\it match} macrolens solutions for B1152+199 system. The outer boxes correspond to enlarged regions around the two macro images. The black solid line on each image depicts the trend an intrinsically straight source would follow at the position of each image given the macrolens model. The gray solid line is the result of fitting a degree two polynomial to the recovered source positions. Any observational hint of curvature in image A would correspond to intrinsic jet curvature in the source plane, and the apparent straightness of jet A therefore places a constraint on the intrinsic curvature of the jet. To evaluate the impact of any such potential curvature, we fit a degree two polynomial to the recovered source, and run it through the lensing software. The slightly curved, gray solid line superposed on image B stems from this intrinsic source curvature derived from the curvature in image A. The yellow solid lines on each panel show the best--fit degree two polynomial to the observed peaks on each image. One can see that both the black and the gray lines provide an acceptable fit to the yellow curvature in image A, as well as the image B in green (EVN data), while deviating slightly from the image B in orange (VLBA data). 

Macroimages that appear curved along the lens caustic are common in strong lensing situations where the source is an extended object. However, the very small apparent size of the image A jet ($\approx 20$ mas) compared to the macrolens Einstein radius ($\approx 0.75$ arcseconds; see Figure~\ref{figure:macrolens_solution}) implies that very little curvature is expected due to the macrolens across the face of this image. Indeed, an intrinsically straight jet produces a perfectly acceptable fit to image A, and results in an image B jet that is perfectly straight down to scales smaller than the beam size. This is demonstrated in the bottom row of Figure ~\ref{figure:macrolens_solution}.

While the jet in image A appears fairly straight, the data quality also allows for slight upward curvature, which due to the reversed image parity between the images would translate into a downward curvature in image B. In Figure~\ref{figure:macrolens_solution} we have attempted to fit the contour map of image A to a two--degree polynomial with the maximum amount of curvature allowed by the data and translated this into the corresponding prediction for the jet shape in image B. The resulting expectation for the image B jet exhibits no difference between the straight and the curved source (black and gray lines respectively).

In comparison to the analysis by \citet{Metcalf02}, we find that our macrolens model provides a considerably better fit to the overall direction of the image B jet and can fully account for the observations. The apparent curvature in the VLBA data, the alternative form of the EVN dataset, and the GVLBI dataset are discussed in section \ref{subsection:beam_effect}.

\section{Discussion}
\label{section:discussion}
The combination of data sets presented here (spanning $\sim$15 years in time), does not provide enough supporting evidence for any jet curvature in image B of B1152+199. We showed that our best--fit models constrained by image A not only provide correct predictions on positions of corresponding blobs in image B, but also correctly reproduced the position angle of the jet (see Figure \ref{figure:curvature}). Even though well within the uncertainty provided by the size of the beam, the apparent small--scale curvature in the VLBA and gVLBI datasets call for an explanation. 

\subsection{Beam effect}
\label{subsection:beam_effect}
\begin{figure}
\includegraphics[width=0.5\textwidth]{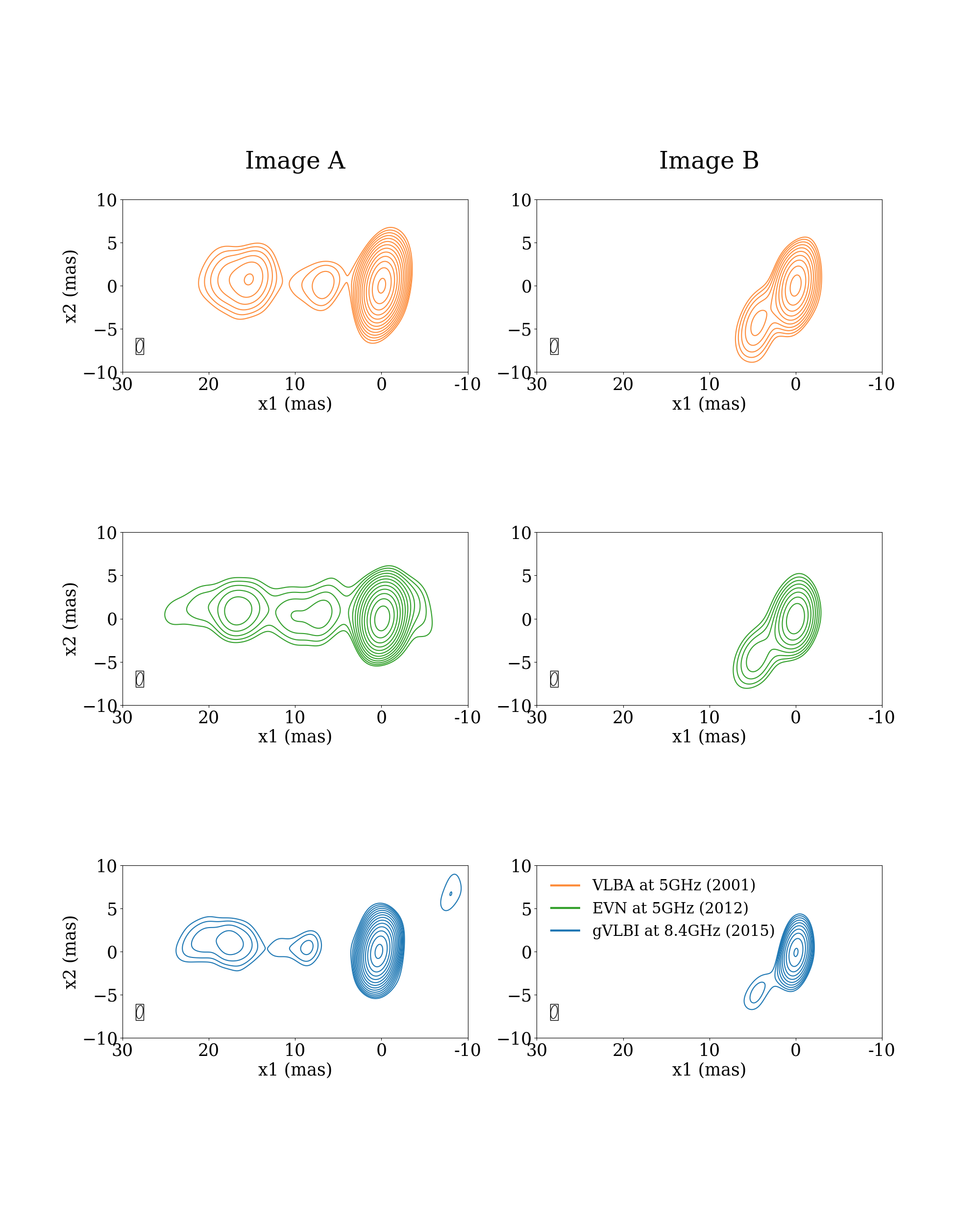}
\caption{The experiment that shows the effect of beam size and ellipticity on the apparent curvature in image B. In this experiment, we convolve three datasets presented in Figure \ref{figure:jet_images} to the same beam to show the effect of beam ellipticity and position angle on the apparent curvature in image B.}
\label{figure:beam_conv} 
\end{figure}

We believe that the most likely explanation for the apparent curvature in image B lies in the effect of the synthesized beam on the actual emissions. Given the elongated shape of the beam, in VLBA 5GHz data and the gVLBI 22GHz as opposed to the more circular beam in the EVN 5GHz data, we performed an experiment in which we convolved both VLBA and EVN data to the same resolution. We used an elongated beam similar to the VLBA BB0133 programme (same orientation) but 30\% larger along both axes with the same position angle (Fig. \ref{figure:beam_conv}). So, by convolving the EVN image with this beam we get an idea of how the EVN structure would look if observed with only the VLBA. Then, we convolve the EVN image to the same resolution. Convolving the VLBA image to this beam will only smooth the image marginally. Indeed, the convolved VLBA image shows the bending almost as clear as the original one. Overall, the convolved VLBA and convolved EVN images look very similar. Based on this observations, we conclude that the bending is indeed a convolution effect. As the EVN image at maximum resolution with a more circular beam shows no bending, and given that this image is much more sensitive (more antennas etc.) than the VLBA image, this could explain the apparent structure. For completeness, we also convolved the 8.4\:GHz image of the B--image to the same resolution. Although slightly lower contours are used here since the emission is weaker, it again resembles well the other two convolved images. 

This experiment suggests that the case for the curvature may be merely an effect of the elongated beam and even the best dataset among those included in this study are unable to robustly determine that the jet in image B is curved.

\section{Conclusions}
\label{section:conclusions}
Using new VLBI data at 5\,GHz (EVN) and 8.4\,GHz (global VLBI) together with the archival VLBA data, carefully calibrated and imaged again, we have revisited the claimed case for curvature and gravitational millilensing in one of the images of the strongly--lensed quasar jet B1152+199 previously discussed by \citet{Rusin02} and \citet{Metcalf02}. Through the analysis presented in this study, we refute the presence of previously claimed curvature in image B within the size of the smallest synthesized beam. Hence, we find no need for an additional secondary lens, and argue that the apparent curvature could be explained as a mere artifact of the synthesized beam.

\section*{Acknowledgements}
EZ acknowledges funding from the Swedish Research Council (project 2011--5349) and the Wenner--Gren Foundations. The European VLBI Network (www.evlbi.org) is a joint facility of independent European, African, Asian, and North American radio astronomy institutes. Scientific results from data presented in this publication are derived from the following EVN project codes:  EJ010 (PI Jackson),  EZ024 (PI Zackrisson), and GA036 (PI Asadi). The National Radio Astronomy Observatory is a facility of the National Science Foundation operated under cooperative agreement by Associated Universities, Inc.
%%%%%%%%%%%%%%%%%%%%%%%%%%%%%%%%%%%%%%%%%%%%%%%%%%



\bibliographystyle{mnras}
\bibliography{refs} % if your bibtex file is called example.bib
%%%%%%%%%%%%%%%%%%%%%%%%%%%%%%%%%%%%%%%%%%%%%%%%%%

% Don't change these lines
\bsp	% typesetting comment
\label{lastpage}
\end{document}